\documentclass[12pt]{article}
\usepackage{latexsym}
\usepackage{amsmath}
\usepackage{amssymb}
\usepackage{epsfig,graphics}
\usepackage{graphicx}
\usepackage{booktabs}
\usepackage{multirow}
\usepackage{caption}
\usepackage{subcaption}
\usepackage{tikz}
\usepackage{fixmath}
\usetikzlibrary{matrix,snakes,arrows,shapes,decorations.pathmorphing,decorations.markings,calc}

\newcommand{\be}{\begin{equation}}
\newcommand{\ee}{\end{equation}}

\begin{document}

\begin{titlepage}

\vspace*{0.6in}
 
\begin{center}
  {\large\bf
    SO(4), SO(3) and SU(2) gauge theories in 2+1 dimensions:\\
    comparing glueball spectra and string tensions}\\
\vspace*{0.75in}
{Michael Teper \\
\vspace*{.25in}
Rudolf Peierls Centre for Theoretical Physics, University of Oxford,\\
1 Keble Road, Oxford OX1 3NP, UK\\
\centerline{and}
All Souls College, University of Oxford,\\
High Street, Oxford OX1 4AL, UK}
\end{center}

\vspace*{0.4in}

\begin{center}
{\bf Abstract}
\end{center}

We improve upon recent calculations of the low-lying `glueball' spectra
of $SO(3)$ and $SO(4)$ lattice gauge theories in 2+1 dimensions, and
compare the resulting continuum extrapolations with $SU(2)$.
We find that these are reasonably consistent,
as are the $SU(2)$ and $SO(4)$ string tensions when these are corrected
for the differing representations of the flux. All this indicates that the
different global properties of these groups do not play a significant role in
the low-lying physics.

\vspace*{0.95in}

\leftline{{\it E-mail:} mike.teper@physics.ox.ac.uk}

\end{titlepage}

\setcounter{page}{1}
\newpage
\pagestyle{plain}

\tableofcontents

\section{Introduction}
\label{section_intro}

Certain $SU(N)$ and $SO(N^\prime)$ groups share the same Lie algebra,
although their global structure differs: that is to say, $SO(3)$ and $SU(2)$,
$SO(4)$ and $SU(2)\times SU(2)$, $SO(6)$ and $SU(4)$. It is
interesting to ask which if any of the physical properties of the
corresponding gauge theories are sensitive to these global differences.
A recent paper
\cite{RLMT_GK}
has addressed this question for the lightest glueball masses and the string
tension in the context of a calculation of $SO(N)$ gauge theories for all $N$.
Unfortunately most of the masses calculated in $SO(N)$ at small values of
$N$ turned out to be subject to large systematic errors, making the
desired comparison of limited value. In this paper we will improve
substantially upon the $SO(4)$ calculations, and significantly upon the
$SO(3)$ ones, allowing us to make a more useful comparison than that
provided in
\cite{RLMT_GK}.
Since the calculations in
\cite{RLMT_GK}
were for gauge theories in $2+1$ dimensions, so are the calculations
in the present paper.

Since this work represents a direct continuation of (some of) the work in
\cite{RLMT_GK}
we will not repeat the discussion in that paper and will simply focus
on the way that our calculations differ.
For a detailed discussion of methods etc. we direct the reader to
\cite{RLMT_GK}.

As an aside we remark that another paper 
\cite{RLMT_Tc}
has compared the deconfining temperature in $SO(4)$ and $SO(6)$ gauge
theories against existing calculations for $SU(2)$ and $SU(4)$ respectively.
The continuum values, expressed in units of the string tension and
corrected for the differing representations, agree within two standard
deviations.

\section{Lattice preliminaries}
\label{section_lattice}

We replace continuum (Euclidean) space-time by a cubic lattice with a lattice
spacing labelled by $a$. The lattice is of finite size,
$L_x\times L_y \times L_t$ in lattice units, with periodic boundary conditions.
The variables are $N\times N$ $SO(N)$ matrices, $U_l$, living on the lattice
links $l$ that join neighbouring sites. The Euclidean path integral is 
$Z=\int {\cal{D}}U \exp\{-S[U]\}$ where ${\cal{D}}U$ is the Haar masure. We 
use the standard plaquette action,
\begin{equation}
S = \beta \sum_p \left\{1-\frac{1}{N} Tr U_p\right\}  \quad ; \quad \beta=\frac{2N}{ag^2}
\label{eqn_S}
\end{equation}
where $U_p$ is the ordered product of link matrices around the plaquette $p$.
As a convenient shorthand, we shall use $u_p \equiv \frac{1}{N} Tr U_p$.
We have written $\beta=2N/ag^2$, but strictly speaking this $ag^2$ is just one 
possible definition of the dimensionless coupling on the length scale $a$.
So if we were to be more precise we would write  $\beta=2N/ag_p^2$ with 
$ag^2_p \stackrel{a\to 0}{=}ag^2 + ca^2g^4 +... \stackrel{a\to 0}{\to} ag^2 $ where $g^2$
is the coupling in the continuum limit.

The purpose of the lattice calculations is to calculate the physics of
the corresponding continuum gauge theories. Since $ag^2$, the dimensionless
running coupling on the scale of the lattice spacing,
vanishes as $a\to 0$ one can determine the leading lattice corrections
just as one does for asymptotically free gauge theories in $3+1$
dimensions. So we know that ratios of two physical masses satisfy
\begin{equation}
  \left.\frac{am_1}{am_2}\right|_a \equiv \left.\frac{m_1}{m_2}\right|_a
  \stackrel{a\to 0}{=}
  \left.\frac{m_1}{m_2}\right|_{a=0} + c(am_3)^2 + O(a^4)
\label{eqn_mM_cont}
\end{equation}
where $am_3$ is some other mass (which may equal $am_1$ or $am_2$) and $c$
is some constant (that depends of course on $m_1, m_2, m_3$).
We can then use this to extrapolate our results to the
continuum limit. We can also extrapolate a lattice mass $am$ expressed in
units of a lattice coupling $ag^2$ using
\begin{equation}
  \left.\frac{am}{ag^2}\right|_a =\left.\frac{m}{g^2}\right|_a 
  \stackrel{a\to 0}{=}
  \left.\frac{m}{g^2}\right|_{a=0} + c ag^2 + O(a^2).
\label{eqn_mg_cont}
\end{equation}
Here the leading correction is $O(a)$ rather than $O(a^2)$ because
the running dimensionless coupling is not a physical quantity even
though the continuum coupling is. The most straightforward
definition of a coupling is $\beta=2N/ag^2$ as in eqn(\ref{eqn_S}).
However we prefer to use the mean-field improved coupling
\begin{equation}
  \beta_I = \beta \langle u_p \rangle \equiv \frac{2N}{ag_I^2}
\label{eqn_betaI}
\end{equation}
since there are some theoretical arguments
\cite{betaI}
that the lattice corrections are smaller using this coupling, as
appears to be borne out in practice for $SU(N)$ gauge theories
\cite{MT_98}.

The energy of a state can be calculated from the time dependence of a
correlation function
\begin{equation}
\langle \phi^\dagger(t) \phi(0) \rangle 
= \sum_n |\langle vac|\phi^\dagger|n \rangle|^2 e^{-E_n t} 
\stackrel{t\to\infty}{\propto} e^{-Mt} 
\label{eqn_M}
\end{equation}
where $M$ is the mass of the lightest state with the quantum numbers
of the operator $\phi$. The operator $\phi$ will typically be the trace
of the product of $SO(N)$ link matrices around some closed path.
To have good overlaps onto the
desired states, so that one can evaluate masses at values of $t$
where the signal has not yet disappeared into the statistical noise,
one needs to use blocked and smeared operators. (For more details see 
\cite{RLMT_GK}.)
This method can be generalised to a variational calculation over
a basis of operators, thus allowing for the efficient calculation
of the energies of excited states 
\cite{RLMT_GK}.
The blocking/smearing involves a parameter that determines the
weighting of the direct path between two sites and the `staples'
that constitute the smearing. In the $SO(N)$ calculations of
\cite{RLMT_GK}
one used the same value for this smearing parameter as the one that
had previously been used in $SU(N)$ calculations
\cite{AAMT_GK,BLMTUW_SUN}.
In this paper we have attempted to optimise the value of this
parameter for $SO(3)$, $SO(4)$ and $SO(6)$ by performing calculations
with various values of the parameter. We find that for $SO(3)$
the best parameter should be chosen to be about 2.5 times the value
used in
\cite{RLMT_GK}
and  for $SO(3)$ about 1.7 times. For $SO(6)$ we find there is much less to
be gained and for that reason we do not attempt to redo the $SO(6)$
calculations in this paper.

To illustrate the improvement we obtain with our operators in this paper,
it is useful to define an effective mass $M_{eff}(t)$ by
\begin{equation}
\frac{\langle\Psi^\dagger(t+a) \Psi(0)\rangle}{\langle\Psi^\dagger(t) \Psi(0)\rangle}
=
e^{-aM_{eff}(t)}.
\label{eqn_Eeff}
\end{equation}
Comparing eqn(\ref{eqn_M}) and eqn(\ref{eqn_Eeff}) it is clear that
what we want is for $M_{eff}(t)$ to approach its asymptotic value
as quickly as possible as $t$ increases, i.e. that the normalised overlap of 
our operator $\Psi$ on the desired ground state should be as close to unity
as possible. We display in Fig.~\ref{fig_Meff_Jp_so4MTRL} the effective
masses for a sample of glueball states ranging from the lightest to the heaviest
in one of our $SO(4)$ calculations. The operators used are the best variational
operators using either the basis of this paper (solid points) or the basis
(open points) of
\cite{RLMT_GK}.
The horizontal lines are our best mass estimates obtained from the apparent
large-$t$ limits of the effective masses using the new operators. 
It is clear that for all the states, apart from the very lightest glueball, the new
operators provide a much more credible estimate of the large-$t$ plateau of
$M_{eff}(t)$ which corresponds to the ground state mass. It is nonetheless also
clear that even with these new operators identifying an effective mass plateau
for the very heaviest states is significantly uncertain. Since the lattice
used in Fig.~\ref{fig_Meff_Jp_so4MTRL} has the second smallest lattice spacing
of our $SO(4)$ calculation, this caveat certainly applies to our eventual
continuum extrapolations.

In  Fig.~\ref{fig_Meff_Jp_so3p1.70p0.75} we provide a similar effective
mass comparison for our $SO(3)$ calculation at the smallest lattice
spacing. While the new calculation is clearly better than that in
\cite{RLMT_GK}
the improvement is less striking than for $SO(4)$. The reason for this
is presumably that for $SO(3)$ the only difference with the calculations of
\cite{RLMT_GK}
is in the choice of the smearing parameter, while for $SO(4)$ the operator
basis used in
\cite{RLMT_GK}
was in addition significantly smaller. It is also clear from a comparison of
Fig.~\ref{fig_Meff_Jp_so3p1.70p0.75} and Fig.~\ref{fig_Meff_Jp_so4MTRL}
that the $SO(4)$ calculations have better overlaps than the $SO(3)$
ones and hence lead to more reliable mass estimates. This is something
already noted in the study of $SO(N)$ gauge theories in
\cite{RLMT_GK}:
as we increase $N$ the overlaps improve. Indeed for $SO(6)$ we found
that changing the smearing parameter did not lead to an improvement
that was large enough to motivate a new calculation.

\begin{figure}[htb]
\begin	{center}
\leavevmode
\input	{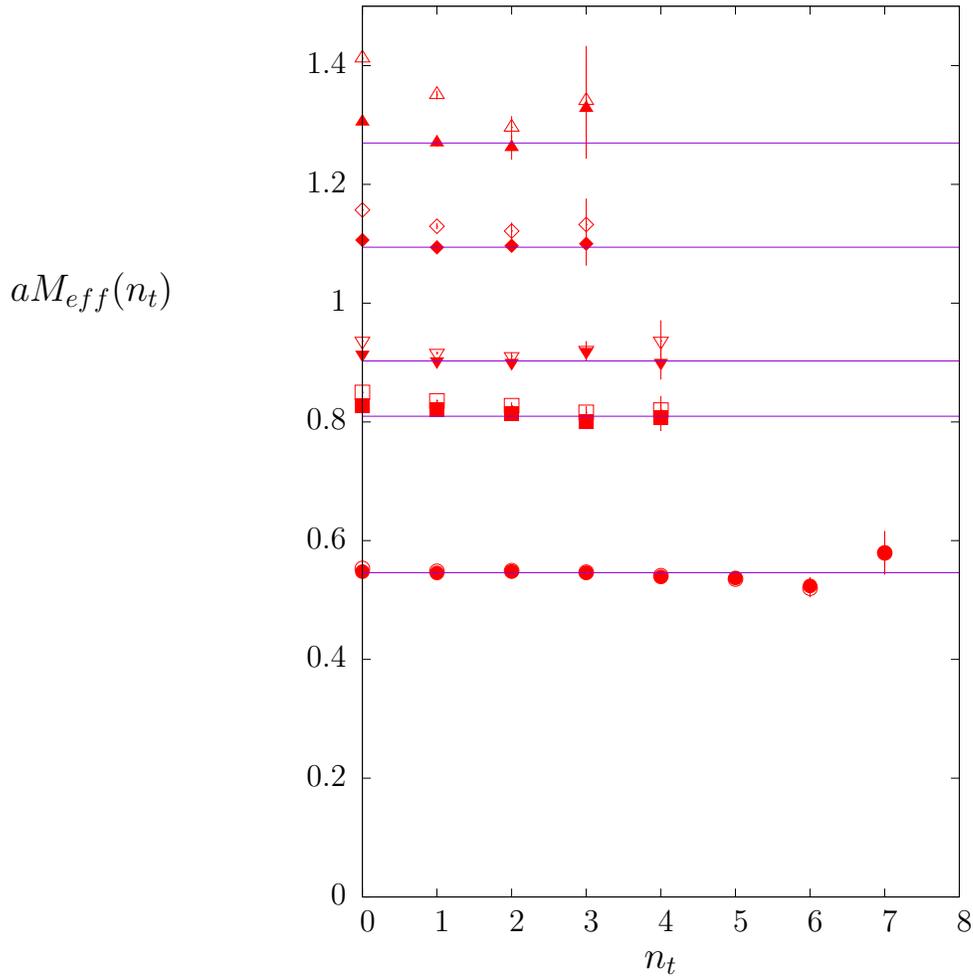}
\end	{center}
\caption{Some glueball effective masses in $SU(4)$ at $\beta=51.0$.
  Solid points are from operators in fundamental representation and open
  points in $k=2$ antisymmetric representation. 
  Glueballs shown are lightest $J^P=0^+$ (circles), first excited $0^+$ (boxes),
  lightest $2^+$ (inverted triangles), first excited $2^+$ (diamonds),
  and lightest $1^+$ (triangles).
  Horizontal lines are the corresponding asymptotic mass estimates using
  operators in fundamental representation.}
\label{fig_Meff_Jp_su4f2A}
\end{figure}

A calculation in
\cite{RLMT_GK}
that helped elucidate the origins of the poor overlaps in
$SO(3)$ was the comparison of the effective mass plots one obtains in $SU(2)$ 
using glueball operators where the trace of the closed loop is taken
in the adjoint and fundamental representations respectively. One found
\cite{RLMT_GK}
that using the adjoint in $SU(2)$ led to poor overlaps similar
to those seen using the fundamental
in $SO(3)$. Since these representations correspond to each other, this
demonstrates that this problem is not peculiar to $SO(N)$ gauge theories.
As an addendum to that we show in Fig.~\ref{fig_Meff_Jp_su4f2A} a similar
comparison for $SU(4)$, but this time using operators in the fundamental
and $k=2$ (antisymmetric) representations, since the latter
corresponds to the fundamental of $SO(6)$. We see that the overlaps of the
$k=2$ operators are visibly worse although the effect is modest except for
the heaviest states. The $SO(6)$ overlaps are quite similar to this.

\section{SO(4) and SU(2)$\times$SU(2)}
\label{section_so4}
\vspace*{0.2cm}

We begin with $SO(4)$ and $SU(2)\times SU(2)$. If the physics of
these two gauge groups is the same then the physics of $SO(4)$ is
just that of two non-interacting
$SU(2)$ groups. That is to say, the single-particle mass spectrum of
$SO(4)$ should be identical to that of $SU(2)$, and taking into
account the differences in the fundamental representations, the
string tensions should differ by a factor of two
\cite{RLMT_GK}
\begin{equation}
\left.\sigma\right|_{so4} = 2\left.\sigma\right|_{su2}
\label{eqn_Kso4}
\end{equation}
and the couplings also by a fctor of two  
\cite{RLMT_GK}
\begin{equation}
\left.g^2\right|_{so4} = 2 \left.g^2_f\right|_{su2}.
\label{eqn_gso4}
\end{equation}

We start with the string tension. In Table~\ref{table_string_so4} we
list our calculated masses of the flux loops winding around the spatial torus
at various bare couplings. From these we can extract the listed values of
the string tension, as explained in
\cite{RLMT_GK}
and we can then extract a continuum value for $\surd\sigma/g^2$ with the fit
\begin{equation}
  \left.\frac{\surd\sigma}{g_I^2N}\right|_{so4} = 0.05889(41) - 0.00048(18)ag_I^2N
  \quad\quad , \, \chi^2/n_{dof}=1.39
\label{eqn_k_so4}
\end{equation}
Taking the above continuum value of $\surd\sigma/g^2$ and
imposing the relations eqns(\ref{eqn_Kso4},\ref{eqn_gso4}) we obtain
a predicted value for $SU(2)$
\begin{equation}
  \left.\frac{\surd\sigma}{g^2}\right|_{so4} = 0.2356(16)
  \Longrightarrow
  \left.\frac{\surd\sigma}{g^2}\right|_{su2} = 0.3332(23).
\label{eqn_k_su2so4}
\end{equation}
This is to be compared to the directly calculated $SU(2)$ value
of $\surd\sigma/{g^2}|_{su2} \simeq 0.3349(3)$, as given in
Table B1 of
\cite{AAMT_GK}.
These values are in agreement within errors.
We can perform the same analysis for the mass gap using the
values listed in Table~\ref{table_glueball_so4}. We obtain
the continuum fit
\begin{equation}
  \left.\frac{m_{0^+}}{g_I^2N}\right|_{so4} = 0.1957(25) - 0.00084(100)ag_I^2N
  \quad\quad , \, \chi^2/n_{dof}=0.48
\label{eqn_mg_so4}
\end{equation}%
from which we obtain a prediction for $SU(2)$
\begin{equation}
  \left.\frac{m_{0^+}}{g^2}\right|_{so4} = 0.783(10)
  \Longrightarrow
  \left.\frac{m_{0^+}}{g^2}\right|_{su2} = 1.566(20).
\label{eqn_m_su2so4}
\end{equation}
This is to be compared to the directly calculated $SU(2)$ value
of $m_{0^+}/{g^2}|_{su2} \simeq 1.586(2)$, as given in
Table B1 of
\cite{AAMT_GK}.
These values are again in agreement within errors. Thus as far as
the mass gap and fundamental string tension are concerned $SO(4)$ is just
like $SU(2)\times SU(2)$, at least within our percent level errors.

\begin{figure}[htb]
\begin	{center}
\leavevmode
\input	{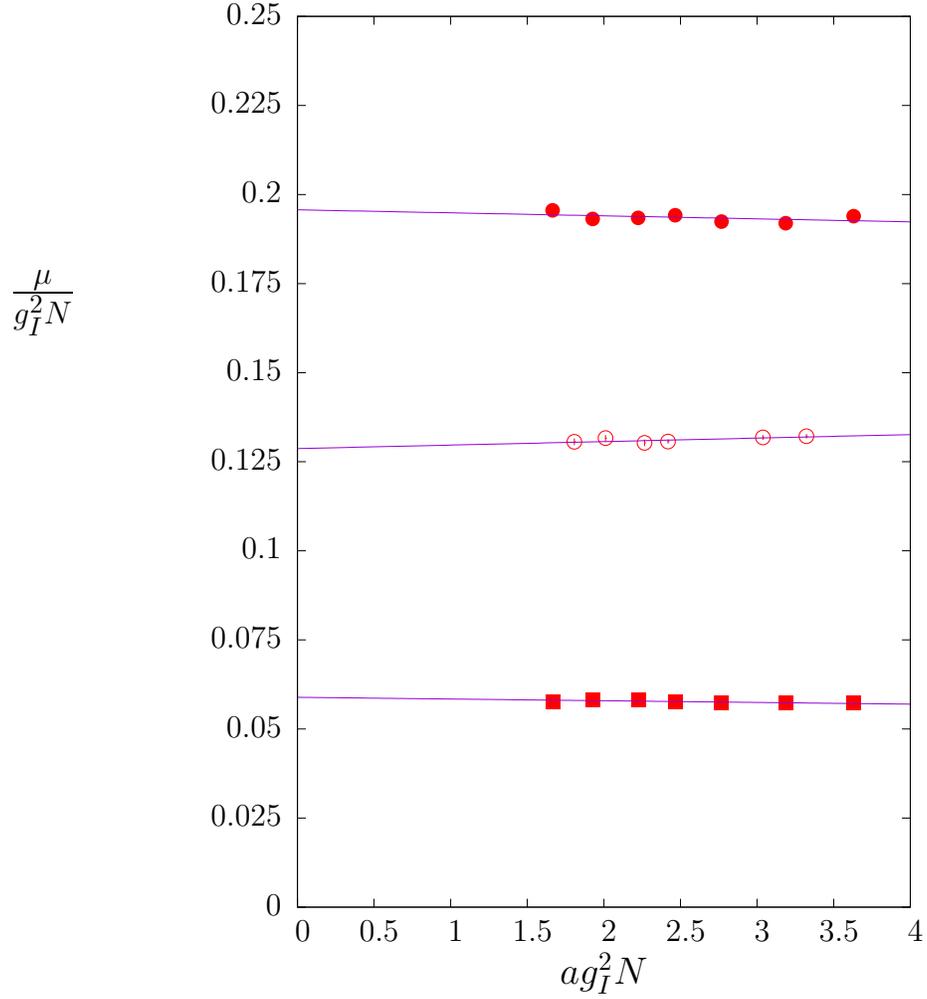}
\end	{center}
\caption{Some continuum masses in units of the lattice 't Hooft coupling,
  with linear extrapolations to the continuum limit : for $\mu = M_{0^+}$ in
  $SO(4)$ ($\bullet$) and in $SO(3)$ ($\circ$), and for $\mu=\surd\sigma_f$
  in $SO(4)$ ($\blacksquare$).}
\label{fig_Mg_cont_soNMT}
\end{figure}

As an aside it is useful to look at the actual lattice data
on a plot versus $ag^2_I$.
This we do for $m_{0^+}/g^2$ and $\surd\sigma/g^2$ in
Fig.~\ref{fig_Mg_cont_soNMT} where we also show our continuum
extrapolations. We note that the best fits shown have very small
slopes: that is to say, these dimensionless
ratios are consistent with having almost no lattice corrections.
On the other hand we can see in Fig.~\ref{fig_Mg_cont_soNMT} that
in units of $ag^2$ the range of our calculated values is not
large, and in fact hardly larger than the range over which we
need to extrapolate to the continuum. This is uncomfortable: a long
extrapolation may conceal significant systematic errors.
To avoid this we can instead look 
at physical mass ratios where the leading correction is
$O(a^2\mu^2)$ (with $\mu$ some reasonable mass scale such as
$m_{0^+}$ or $\surd\sigma$) in which case our calculated values
extend to a region much closer to the continuum limit, as we shall
now see.

\begin{figure}[htb]
\begin	{center}
\leavevmode
\input	{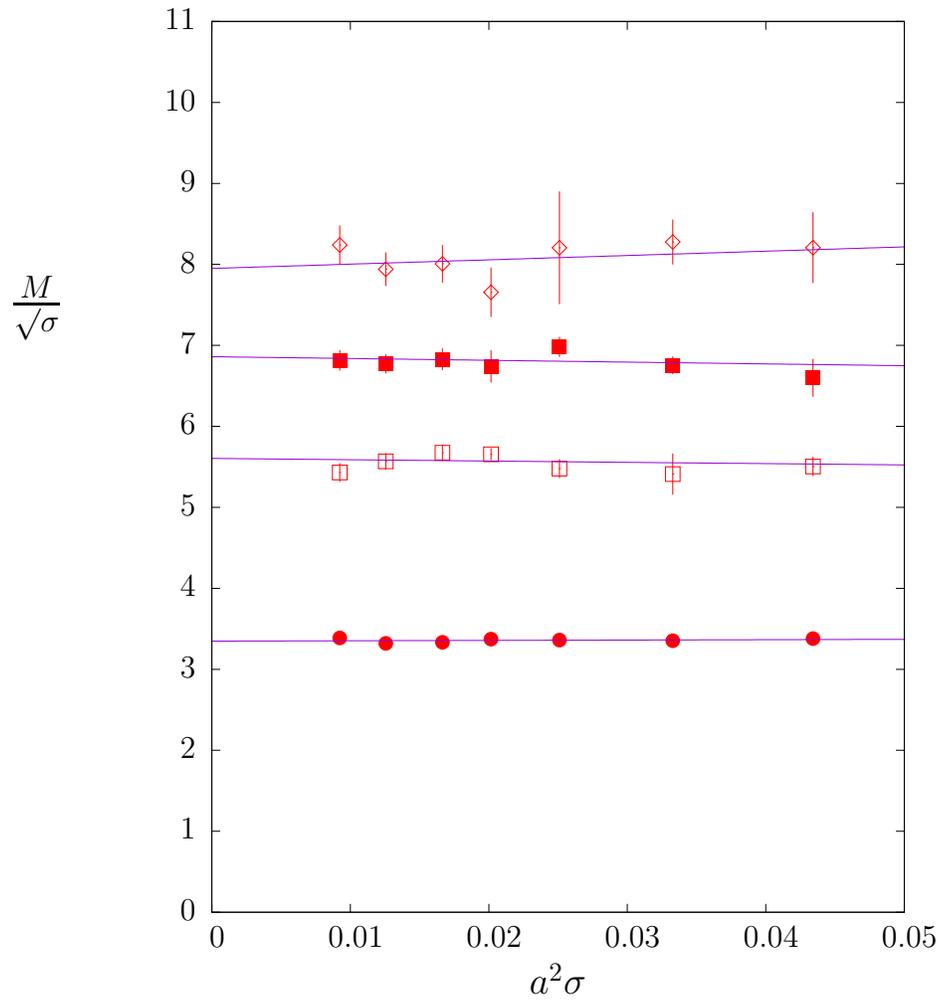}
\end	{center}
\caption{Some masses in units of the string tension in $SO(4)$,
  with linear extrapolations to the continuum limit : for $M = M_{0^+}$ 
  ($\bullet$), $M = M_{2^-}$ ($\circ$), $M = M_{2^{-\star}}$ ($\blacksquare$),
  and $M = M_{1^{-}}$ ($\diamond$).}
\label{fig_MK_cont_so4MT}
\end{figure}

In Table~\ref{table_glueball_so4} we list our best estimates for
the masses of the same glueball states as calculated in
\cite{RLMT_GK}.
As can be seen from the examples of effective masses in
Fig.\ref{fig_Meff_Jp_so4MTRL} the heavier the glueball the
less reliable is our estimate of its mass. We perform
continuum extrapolations of the dimensionless ratio
$m/\surd\sigma$ using the string tensions listed in
Table~\ref{table_string_so4}. In general we obtain acceptable
fits to our masses with just a lowest order $O(a^2\sigma)$
lattice correction and we use these fits to estimate the continuum
mass ratios listed in Table~\ref{table_glueball_SO4vsSU2}, where
we also give the $\chi^2$ per degree of freedom of each fit.
A selection of continuum extrapolations are shown in
Fig.~\ref{fig_MK_cont_so4MT}. Although the displayed best fits have
small but non-zero slopes, in fact the slopes are consistent with zero
within errors, i.e. consistent with negligible lattice corrections.
This is in fact the case for our continuum fits to all the
states in Table~\ref{table_glueball_SO4vsSU2}. Moreover
here the range over which we have calculated masses extends
much closer to the continuum limit, largely eliminating any worry
about associated systemetic errors.

In Table~\ref{table_glueball_SO4vsSU2} we also list the corresponding
mass ratios in $SU(2)$. We see that we have good agreement,
within $1\sigma$, between the $SO(4)$ and $SU(2)$ values
for the $0^+, 0^{+\star}, 2^-$ and $0^-$ glueballs, and acceptable
agreement, i.e. within about $2\sigma$, for all the rest
except for the $2^+$, the  $2^{-\star}$ (both about $3\sigma$)
and the $1^+$ (about $4\sigma$). The fact that these larger
discrepancies are all in one direction, i.e. the $SO(4)$ ratios
larger than the $SU(2)$ ones, is consistent with the fact
that for heavier states our identification of an effective mass
plateau is weak, and by extracting a mass too early along the
correlator we overestimate the mass. Nonetheless we see that
the bulk of the extensive disagreement that was noted in
\cite{RLMT_GK}
has disappeared in our new calculations.

\section{SO(3) and SU(2)}
\label{section_so3}
\vspace*{0.2cm}

We turn now to $SO(3)$. A difference between $SO(3)$ and $SO(4)$ is that
the fundamental flux tube in $SO(3)$ is visibly unstable 
\cite{RLMT_GK},
which is no surprise given that it corresponds to the adjoint
flux tube of $SU(2)$. So extracting a string tension in $SO(3)$
is delicate and the result will not be very precise.  Since
we are interested in comparisons with $SU(2)$ that are  precise
enough to be significant, we will not consider the $SO(3)$ string
tension any further in this paper.

We begin with the lightest glueball. In Table~\ref{table_massgap_so3} we
list our calculated values of the mass gap at various bare couplings. 
From these we can extract a continuum value for $m_{0^+}/g^2$ with the fit
\begin{equation}
  \left.\frac{m_{0^+}}{g_I^2N}\right|_{so3} = 0.1287(16) + 0.00098(41)ag_I^2N
  \quad\quad , \, \chi^2/n_{dof}=0.6
\label{eqn_mg_so3}
\end{equation}
as displayed in Fig.~\ref{fig_Mg_cont_soNMT}.
Now for $SU(2)$ one finds $m_{0^+}/g^2|_{su2} = 1.5860(22)$. (See Table B1 of
\cite{AAMT_GK}.)
So if we assume that the mass gap is identical in $SU(2)$ and $SO(3)$, the
above values of $m_{0^+}/g^2$ in $SO(3)$ and $SU(2)$ imply the following
relationship between the couplings:
\begin{equation}
  \frac{g^2_{so3}}{g^2_{su2}} = 4.108(51).
\label{eqn_g_su2so3}
\end{equation}
This should be compared to the theoretically  expected value of 4
\cite{RLMT_GK}.
The agreement is within an acceptable $2\sigma$.
Alternatively we could assume this factor of 4
and then obtain a prediction for  $m_{0^+}/g^2$ in $SU(2)$ which would
be within  about $2\sigma$ of the known value.

The continuum extrapolations in units of $ag^2$ suffer from their distance
from the continuum limit, as is evident from Fig.~\ref{fig_Mg_cont_soNMT}.
As in the case of $SO(4)$, a solution is to extrapolate 
ratios of physical masses.
Since we do not have a usefully precise string tension to use
in forming our dimensionless mass ratios we need to use some other
quantity and we choose to use the mass gap since, being
the lightest mass, it should also be the one that is calculated most
accurately. So we extrapolate the ratio $m_G/m_{0^+}$ with an
$O(a^2m^2_{0^+})$ lattice correction for the various glueballs $G$,
whose calculated masses are listed in Table~\ref{table_glueball_so3}.
The resulting continuum values of these ratios
are listed in Table~\ref{table_glueball_SONvsSUM} together
with those for $SU(2)$ (and also, to be complete, those for $SO(4)$).
The $SU(2)$ and $SO(3)$ mass ratios agree within $\sim 2\sigma$ for the
$0^{+\star}$, the $0^{+\star\star\star}$, and the $0^{+\star\star\star\star}$,
and also for the $2^{+}$, the $2^{-}$, the $2^{+\star}$, the $2^{-\star}$. 
Of the remaining four glueballs only the  $0^{+\star\star}$ differs from 
the $SU(2)$ value by more
than $\sim 3\sigma$. We note that just as in the case of $SO(4)$
all the significant differences are
due to the $SO(3)$ mass estimates being larger than the $SU(2)$ ones,
which is consistent with the systematic error incurred when
extracting a mass too early in the correlator. The examples
of effective mass plots in Fig.~\ref{fig_Meff_Jp_so3p1.70p0.75}
show that while our new mass calculations involve operators that
have a significantly better projection onto the various glueball
states than the operators used in
\cite{RLMT_GK},
they are not yet adequate for identifying unambiguously the effective
mass plateaux of the heavier glueballs such as the $0^-$ and $1^\pm$.

\section{Conclusions}
\label{section_concl}

Our main motivation for the calculations in this paper was the apparent
disagreement, observed in
\cite{RLMT_GK},
between much of the low-lying spectrum of the $SO(4)$ gauge theory
and that of $SU(2)$, and, to a lesser extent, that of $SO(3)$ as well.
It was argued in
\cite{RLMT_GK}
that this was probably due to the relatively poor overlap of the
operators onto the corresponding glueball states. In this paper
we improved the operators used and found much better agreement
between the spectra of $SO(4)$, $SO(3)$ and $SU(2)$ gauge theories.
The improvement has been greatest for $SO(4)$ where, for the lightest
and most reliably calculated single particle states, we now have agreement
with $SU(2)$ and hence with the $SU(2)\times SU(2)$ theory with
which  $SO(4)$ shares its Lie algebra. And similarly for $SO(3)$
and $SU(2)$. We did not attempt to repeat the comparison
between $SO(6)$ and $SU(4)$ because our operator improvement proved
very modest for $SO(6)$ and in any case there was already reasonable
agreement in
\cite{RLMT_GK}
between the low-lying spectra of $SO(6)$ and $SU(4)$, except for
a couple of scalar glueball excited states. So in summary all this points
to the conclusion that the differing global properties
of these $SO(N)$ and $SU(N^\prime)$ groups play no significant role in the
low-lying physics of the corresponding gauge theories.

\section*{Acknowledgements}

The numerical computations were carried out on the computing cluster
in Oxford Theoretical Physics.

\clearpage

\begin{table}[h]
\centering
\begin{tabular}{ |ccccc|c|}
  \hline
  	$L_s^2L_t$ 	& $\beta$ & $\tfrac{1}{N}\text{tr}(U_p)$ & $am_{P}$ & $a\surd\sigma$& $\surd\sigma/g^2N$\\
  \hline
        $22^232$ 	&11.0	&0.8013472(21)	&0.931(16)	&0.2084(18)  &0.05740(48)\\
	$26^236$	&12.2	&0.8229531(4)	&0.8449(92)	&0.1824(10)  &0.05724(30)\\
	$30^240$	&13.7	&0.8440177(4)	&0.7354(74)	&0.1584(8)   &0.05725(28)\\
	$34^244$	&15.1	&0.8595470(3)	&0.6699(63)	&0.1420(7)   &0.05759(26)\\
	$38^248$	&16.5	&0.8722327(2)	&0.6186(40)	&0.1290(4)   &0.05803(18)\\
	$42^252$	&18.7	&0.8880800(2)	&0.5150(33)	&0.1121(4)   &0.05817(18)\\
  $46^260$	&21.3	&0.9023628(2)	&0.4134(33)	&0.0961(4)   &0.05773(22)\\
  \hline
  --       	&$\infty$& --    	& --    	& --         & 0.05889(41)\\
  \hline
\end{tabular}
	\caption{$SO(4)$ average plaquette values, flux loop masses, and string tensions.}
	\label{table_string_so4}
\end{table}

\begin{table}[h]
\centering
\begin{tabular}{|l|lllllll|}
  \hline
$L_s^2L_t$ 	& $34^242$  & $42^246$  & $46^252$  & $50^256$  & $58^262$ & $66^270$ & $74^280$\\
$\beta$		& 11.0	   & 12.2      & 13.7	   & 15.1      & 16.5     & 18.7     & 21.3 \\
  \hline
$0^+$		& 0.7040(59) & 0.6118(58) & 0.5325(44) & 0.4788(46) & 0.4301(37) & 0.3722(33) & 0.3526(26) \\
$0^{+*}$	& 0.983(30) & 0.902(19) & 0.787(12) & 0.671(36) & 0.588(22) & 0.522(15) & 0.4672(87) \\
$0^{+**}$	& 1.243(19) & 1.125(62) & 0.961(27) & 0.877(25) & 0.784(12) & 0.633(26) & 0.595(20) \\
$0^{+***}$	& 1.438(39) & 1.27(11) & 1.063(42) & 0.892(58) & 0.838(56) & 0.760(20) & 0.662(13) \\
$0^{+****}$	& 1.496(48) & 1.318(26) & 1.047(55) & 0.962(36) & 0.942(24) & 0.792(32) & 0.703(14) \\

$2^{+}$		& 1.150(16) & 1.027(13) & 0.887(26) & 0.795(16) & 0.729(9) & 0.619(12) & 0.549(7) \\
$2^{+*}$	& 1.432(55) & 1.217(23) & 1.070(41) & 0.954(20) & 0.868(18) & 0.751(20) & 0.652(14) \\
$2^{-}$		& 1.147(23) & 0.987(46) & 0.868(18) & 0.803(10) & 0.732(13) & 0.624(11) & 0.522(11) \\
$2^{-*}$	& 1.375(47) & 1.232(19) & 1.106(19) & 0.957(28) & 0.881(17) & 0.759(13) & 0.655(12) \\

$0^{-}$		& 1.519(62) & 1.354(46) & 1.089(62) & 1.021(27) & 0.925(26) & 0.740(55) & 0.689(38) \\

$1^{+}$		& 1.73(11) & 1.52(8) & 1.124(59) & 1.154(54) & 1.028(37) & 0.925(15) &  0.812(25) \\
$1^{-}$		& 1.71(9) & 1.51(5) & 1.30(11) & 1.087(43) & 1.033(30) & 0.890(23) & 0.792(23) \\
  \hline
\end{tabular}
	\caption{$SO(4)$ glueball masses $am_G$.}
	\label{table_glueball_so4}
\end{table}

\begin{table}
\centering
\begin{tabular}{|c||ll|l|}
  \hline
  \multicolumn{4}{|c|}{ $rM_G/\surd\sigma$} \\ \hline
  $J^P$   &  $SO(4)$ & $\chi^2/n_{dof}$ &  $SU(2)$  \\	\hline
  
$0^{+ }$	& 4.735(40) & 0.66  & 4.737(6)  \\
$0^{+*}$	& 6.77(15)  & 1.70  & 6.861(14)  \\
$0^{+**}$	& 8.64(19)  & 0.88  & 8.382(14)  \\
$0^{+***}$	& 9.63(21)  & 0.55  & 9.278(16)  \\
$0^{+****}$	& 10.23(23) & 1.30  & 9.708(21)  \\
$2^{+}$		& 8.06(11)  & 0.48  &  7.762(10)  \\
$2^{-}$         & 7.93(14)  & 0.98  &  7.795(12) \\
$2^{+ *}$       & 9.53(21)  & 0.13 &  9.107(20) \\
$2^{- *}$       & 9.69((18) & 0.61 &  9.123(24) \\
$0^{-}$	       & 9.86(38)  & 0.45 &  9.884(25) \\
$1^{+}$	       & 11.87(33) & 1.84 &  10.553(31) \\
$1^{-}$	       & 11.24(33) & 0.64 &  10.538(28) \\
  \hline
\end{tabular}
\caption{Continuum glueball masses in units of the string tension: the
  spectrum of $SO(4)$ gauge theory versus that of $SU(2)$, with $r=1, \surd 2$
  for  $SU(2), SO(4)$ respectively.}
	\label{table_glueball_SO4vsSU2}
\end{table}

\begin{table}[h]
\centering
\begin{tabular}{ |ccccc|}
  \hline
  	$L_s^2L_t$ 	& $\beta$ & $\tfrac{1}{N}\text{tr}(U_p)$ & $am_{0^+}$ & $m_{0^+}/g^2N$\\
  \hline
        $54^240$ 	&6.5   &0.8335239(13)	& 0.4391(21) & 0.1322(7) \\
	$62^248$	&7.0   &0.84655214(20)  & 0.4004(21) & 0.1318(7) \\
	$74^260$	&8.5   &0.87556397(12)  & 0.3161(14) & 0.1307(6) \\
	$82^264$	&9.0   &0.88291502(10)  & 0.2952(21) & 0.1303(10) \\
	$90^270$	&10.0  &0.89526817(7)   & 0.2647(17) & 0.1317(9) \\
	$100^280$	&11.0  &0.90524861(6)   & 0.2361(17) & 0.1306(10) \\
  \hline
  --       	&$\infty$& --    	& --        & 0.1287(16)  \\
  \hline
\end{tabular}
\caption{$SO(3)$ average plaquette values, the mass gap
  and its continuum extrapolation.}
	\label{table_massgap_so3}
\end{table}

\begin{table}
\centering
\begin{tabular}{|l|llllll|}
  \hline
$L_s^2L_t$ 	&$54^240$  &$62^248$  &$74^260$  &$82^264$  &$90^270$  &$100^280$   \\
$\beta$		&6.5	   &7.0	     &8.5	&9.0	   &10.0       &11.0 \\
  \hline
$0^+$		& 0.4391(21) & 0.4004(21)  & 0.3161(14)  & 0.2952(21)  & 0.2647(17)  & 0.2361(17)  \\
$0^{+*}$	& 0.6461(76) & 0.5786(44)  & 0.4637(37)  & 0.4384(38)  & 0.3890(40)  & 0.3438(38)  \\
$0^{+**}$	& 0.766(13)  & 0.732(9)    & 0.576(11)   & 0.542(8)    & 0.4908(49)  & 0.4287(86)  \\
$0^{+***}$	& 0.848(18)  & 0.773(13)   & 0.617(15)   & 0.601(10)   & 0.533(13)   & 0.4655(93)  \\
$0^{+****}$	& 0.883(21)  & 0.833(20)   & 0.642(14)   & 0.592(23)   & 0.559(14)   & 0.497(17)   \\

$2^{+}$		& 0.707(18)  & 0.645(17)  & 0.520(6)   & 0.4890(72)  & 0.4320(51)  & 0.3875(52)  \\
$2^{+*}$	& 0.882(22)  & 0.790(15)  & 0.6364(67) & 0.581(12)   & 0.5218(67)  & 0.4725(55)  \\
$2^{-}$		& 0.726(10)  & 0.661(9)   & 0.5200(58) & 0.4945(49)  & 0.4351(49)  & 0.3945(42)  \\
$2^{-*}$	& 0.872(25)  & 0.758(32)  & 0.626(12)  & 0.5871(86)  & 0.507(10)   & 0.4684(76)  \\

$0^{-}$		& 0.920(22)  & 0.858(15)  & 0.7086(89) & 0.651(12)  & 0.554(15)  & 0.476(18)  \\

$1^{+}$		& 0.944(48)  & 0.937(23)  & 0.787(27)  & 0.704(19)  & 0.595(21)  & 0.556(8)  \\
$1^{-}$		& 1.011(43)  & 0.932(27)  & 0.753(27)  & 0.738(22)  & 0.630(25)  & 0.546(13)  \\
  \hline
\end{tabular}
	\caption{$SO(3)$ glueball masses $am_G$.}
	\label{table_glueball_so3}
\end{table}

\begin{table}
\centering
\begin{tabular}{|c||ll|ll|l|}
  \hline
  \multicolumn{6}{|c|}{ $M_G/M_{0^+}$}  \\ \hline
  $J^P$ 	& $SO(3)$ & $\chi^2/n_{dof}$ & $SO(4)$ & $\chi^2/n_{dof}$ & $SU(2)$ \\  \hline
$0^{+*}$	& 1.476(18) & 0.89 & 1.424(32) & 1.37 & 1.449(4)   \\
$0^{+**}$	& 1.882(30) & 1.01 & 1.832(40) & 0.66 & 1.770(4)   \\
$0^{+***}$	& 2.042(43) & 0.76 & 2.024(47) & 0.48 & 1.959(4)   \\
$0^{+****}$	& 2.105(60) & 0.61 & 2.149(51) & 1.29 & 2.050(5)   \\
$2^{+}$		& 1.659(29) & 0.25 & 1.700(24) & 0.33 & 1.639(3)   \\
$2^{-}$         & 1.665(22) & 0.47 & 1.675(30) & 1.19 & 1.646(3)   \\
$2^{+ *}$       & 1.991(34) & 0.54 & 2.009(45) & 0.10 &  1.923(5)  \\
$2^{- *}$       & 1.969(43) & 0.85 & 2.050(38) & 0.65 &  1.926(6)  \\
$0^{-}$	       & 2.219(53) & 3.46 & 2.085(80) & 0.43 &  2.087(6)  \\
$1^{+}$	       & 2.391(57) & 1.78 & 2.515(80) & 1.81 &  2.228(7)  \\  
$1^{-}$	       & 2.393(74) & 1.12 & 2.378(70) & 0.56 &  2.225(7)  \\  \hline
\end{tabular}
\caption{Continuum glueball masses in units of the mass gap: comparing  the
  spectra of $SO(3)$ and $SO(4)$ gauge theories with that of $SU(2)$.}
	\label{table_glueball_SONvsSUM}
\end{table}

\clearpage

\begin{figure}[htb]
\begin	{center}
\leavevmode
\input	{plot_Meff_Jp_so4MTRL.tex}
\end	{center}
\caption{Some effective masses in $SO(4)$ at $\beta=18.7$. Solid points are new
  calculations while open points are from the older calculations in
  \cite{RLMT_GK}. Glueballs shown are lightest $J^P=0^+$ (circles), lightest
  $2^+$ (diamonds), first excited $2^+$ (boxes), and lightest $1^+$ (triangles).
  Horizontal lines are our corresponding mass estimates as listed in
  Table~\ref{table_glueball_so4}.}
\label{fig_Meff_Jp_so4MTRL}
\end{figure}

\begin{figure}[htb]
\begin	{center}
\leavevmode
\input	{plot_Meff_Jp_so3p1.70p0.75.tex}
\end	{center}
\caption{Some effective masses in $SO(3)$ at $\beta=11.0$. Solid points are new
  calculations while open points are from the older calculations in
  \cite{RLMT_GK}. Glueballs shown are lightest $J^P=0^+$ (circles), lightest
  $2^+$ (diamonds), first excited $2^+$ (boxes), and lightest $1^+$ (triangles).
  Horizontal lines are our corresponding mass estimates as listed in
  Table~\ref{table_glueball_so3}.}
\label{fig_Meff_Jp_so3p1.70p0.75}
\end{figure}

\end{document}